%% file: asilomar.tex
\newcommand{\CLASSINPUTtoptextmargin}{1.75cm}
\newcommand{\CLASSINPUTbottomtextmargin}{1.75cm}
\newcommand{\CLASSINPUToutersidemargin}{1.7cm}

\documentclass[10pt,twocolumn,conference]{IEEEtran}

\usepackage{amsmath, mathrsfs, mathtools}
\usepackage{amsfonts}
\usepackage{amssymb}
\usepackage{graphicx}
\usepackage{color}
\usepackage{multirow}
\usepackage{graphicx}
\usepackage{stmaryrd}
\usepackage{subfig}

\allowdisplaybreaks

\usepackage{grffile}
\usepackage{amsmath}

\let\labelindent\relax
\usepackage{enumitem}

\newcommand{\subparagraph}{}
\usepackage[compact]{titlesec}
\titlespacing*{\section}{15pt}{1.2\baselineskip}{0.9\baselineskip}

\allowdisplaybreaks

\newcommand{\myhash}{%
  {\settoheight{\dimen0}{C}\kern-.05em\, \resizebox{!}{\dimen0}{\raisebox{\depth}{\#}}}}

\usepackage{caption}

\usepackage[numbers,sort&compress]{natbib}

\usepackage{algorithm}
\usepackage{algpseudocode}
\usepackage{pifont}
\usepackage{varwidth}

\def\gammam{{\boldsymbol{\gamma}}}

\newcommand{\dimPilots}{L} %

\usepackage{multicol}


\input{symbols.tex}

\usepackage{tikz}
\usepackage{pgfplots,tikz-3dplot}
\usetikzlibrary{intersections, pgfplots.fillbetween}
\pgfplotsset{compat=newest}
\usetikzlibrary{patterns}
\usetikzlibrary{positioning}
\usetikzlibrary{datavisualization}
\usetikzlibrary{datavisualization.formats.functions}
\usetikzlibrary{backgrounds}
\usetikzlibrary{shapes,snakes}

\usepgfplotslibrary{external} 
\usepackage{float}
\usepackage{afterpage}
\usepackage{balance}

\def\herm{{\sfH}}

\def\cg{{\clC\clN}} 
\newcommand{\normd}[1]{{\left\vert\kern-0.25ex\left\vert\kern-0.25ex\left\vert #1 
    \right\vert\kern-0.25ex\right\vert\kern-0.25ex\right\vert}}

\input{macros}
\title{Grant-Free Massive Random Access With a Massive MIMO Receiver}
\author{
\IEEEauthorblockN{Alexander Fengler, Saeid Haghighatshoar,  Peter Jung, Giuseppe Caire }
\IEEEauthorblockA{Communications and Information Theory Group,\\
Technische Universit\"{a}t Berlin\\
E-mail: \{fengler, saeid.haghighatshoar, peter.jung, caire\}@tu-berlin.de.}
}

\begin{document}

\maketitle

\begin{abstract}
    We consider the problem of {\em unsourced random access} (U-RA), a grant-free uncoordinated
    form of random access, in a wireless channel with a massive MIMO base station
    equipped with a large number $M$ of antennas and a large number of wireless single-antenna devices (users).
    We consider a block fading channel model where the $M$-dimensional channel vector of each user remains 
    constant over a \textit{coherence block} containing $L$ signal dimensions in time-frequency. 
    In the considered setting, the number of potential users $K_\text{tot}$ 
    is much larger than $\dimPilots$ but at each time slot only $K_a \ll K_\text{tot}$ 
    of them are active. Previous results, based on compressed sensing, require that 
    $K_a\le \dimPilots$,
    which is a bottleneck in massive deployment scenarios such as {\em Internet-of-Things}  
    and U-RA. 
    In the context of activity detection it is known that such a limitation can be overcome when
    the number of base station antennas $M$ is sufficiently large and a covariance based recovery
    algorithm is employed at the receiver. We show that, in the context of U-RA,
    the same concept allows to achieve high spectral efficiencies in the order of $\mathcal{O}(L
    \log L)$, although at an exponentially growing complexity. We show also that a concatenated coding
    scheme can be used to reduce the complexity to an acceptable level while still achieving
    total spectral efficiencies in the order of $\mathcal{O}(L/\log L)$.
\end{abstract}

\begin{keywords}
    Random Access (RA), Internet of Things (IoT), Massive MIMO, Grant-Free RA, Unsourced RA.
\end{keywords}

\section{Introduction}  \label{intro}

One of the paradigms of modern machine-type communications \cite{Tal2012a} consists of a very large
number of devices (here referred to as ``users'') with sporadic data.  Typical examples thereof are
Internet-of-Things (IoT) applications, wireless sensors deployed to monitor smart infrastructure,
and wearable biomedical devices \cite{Has2013}.  In such scenarios, a Base Station (BS) should be
able to collect data from a large number of devices.  However, due to the sporadic nature of the
data generation and communication, allocating some dedicated transmission resource to all users in
the system may be extremely wasteful, especially for short messages.
On a high level, we distinguish between {\em grant-based} and
{\em grant-free} approaches.
In a grant-based protocol the active users are identified and the BS
can then allocate transmission resources to the active users, while in a grant-free protocol the
users transmit their data right away without awaiting the grant approval of the BS.
Interestingly, virtually
any exiting cellular standard in operation today (3G, 4G-LTE, 5G New Radio) makes use of a dedicated
random access channel (or slot), followed by some scheduling and resource allocation for the active
users \cite{Ses2009,Agi2016a}, therefore these schemes can be seen as examples of grant-based
protocols.
In contrast, {\em unsourced random access} (U-RA) is a novel grant-free paradigm proposed in \cite{Pol2017}
and motivated by an IoT scenario where millions of cheap devices have their codebook hardwired at
the moment of production, and are then disseminated into the environment.  In this case,  all users
make use of the very same codebook and the BS must decode the list of transmitted messages
irrespectively of the identity of  of the active users.
\footnote{
    If a user wishes to communicate its ID, it can send it as
    part of the payload. Therefore, in the paradigm of U-RA, if the users make
    use of individually different codebooks, it would be impossible for the BS to know in advance
    which codebook to decode since the identity of the active users is not known a priori. Hence, in
    this context it is in fact essential, and not just a matter of implementation costs, that all users
    utilize the same codebook.
}

The author of \cite{Pol2017} introduced the U-RA model on the
real adder multiple-access channel (MAC) with additive white Gaussian noise (AWGN) and
established quite tight achievability and converse bounds to the minimum energy per
bit over $N_0$ required for reliable communications. It was shown
that classical MAC schemes like TIN or Aloha perform poorly compared to the random coding achievability
bound.
Subsequent works introduced many practical coding schemes
\cite{Ord2017,Cal2018a,Ama2018,Fen2019c,Ama2019,Mar2019,Pra2019}, which successively
decreased the gap to the achievability bound.
In this line of work \cite{Ama2018} proposed a coding scheme based on partitioning the transmission slot into
subslots, and letting each active user send a codeword from a common codebook across the subslots.
The common codebook is obtained by concatenating an outer tree code with an inner “compressed
sensing code”. The inner encoder maps each submessage into one column of a given (real) coding
matrix. The inner decoder must identify which columns of the matrix have been transmitted from the
received noisy superposition. This is a classical sparse support identification problem, well
investigated in the compressed sensing (CS) literature.
The inner decoder produces a sequence of active-submessage-lists
across the subblocks. The task of the outer tree decoder is to “stitch together” the submessages
such that each sequence of submessages is a valid path in the tree.

Most of the work on U-RA focused on the real adder MAC with AWGN. Recently the U-RA model was
extended to the quasi static fading MAC \cite{Kow2019a}, establishing converse and achievability
bounds on $E_b/N_0$. In this work we extend the U-RA model of \cite{Pol2017} to a block-fading MIMO channel,
where the channel coefficients remain constant over \textit{coherence blocks} consisting
of $\dimPilots$ signal dimensions in the time-frequency domain,
and change randomly from block to block according to a stationary ergodic process
\cite{tse2005fundamentals}. We refer to the average power of the active users as
large scale fading coefficients (LSFCs), which are assumed to be deterministic but unknown.
The LSFCs vary between different users because of varying distances and large scale-effects such as
log-normal shadowing. We formulate the U-RA problem as a joint sparse support recovery
problem with multiple measurement vectors (MMVs) \cite{Che2006,Cot2005a,Kim2012}
and leverage recent advances in MIMO activity detection (AD) \cite{AD:isit2018}.

A fundamental limitation when considering AD with a single-antenna BS is that
the required signal dimension $\dimPilots$ to reliably identify a subset of $K_a$ active users among
a set consisting of $K_\text{tot}$ \textit{potentially active} users scales as $\dimPilots=O(K_a
\log(\frac{K_\text{tot}}{K_a}))$, thus, almost linearly with $K_a$. 
The key to overcome the linear scaling of $L$ with $K_a$ consists of considering quadratic
measurements, i.e., sample covariance information.  This observation was already empirically
provided in \cite{Pal2015d} where a ``much better than linear'' regime was experimentally observed
and conjectured to be achievable using LASSO applied to the sample covariance matrix of the
observation. Also in \cite{Pal2015d} the importance of the Khatri-Rao product \cite{Kha1968} for
sparse recovery from MMVs was
established for the first time. However, only a linear scaling law was proved because the
analysis of LASSO based on coherence is too weak.   

In \cite{Fen2019b,Fen2019d} the restricted isometry property (RIP) was shown to hold for
Kathri-Rao product matrices and the RIP was subsequently used to show that a covariance based
non-negative least-squares (NNLS) estimator can identify the activity of up to
$K_a=O(\dimPilots^2/\log^2(\frac{K_\text{tot}}{K_a}))$ active users provided that the number of
antennas at the BS grows faster than $K_a$.
Furthermore, in \cite{AD:isit2018,Fen2019d} we presented an improved algorithm for AD based on
maximum-likelihood (ML) estimation of the activity pattern as an unknown vector with non-negative 
components. The resulting likelihood function minimization is a non-convex
problem, that can be solved (approximately) by iterative componentwise minimization.  
While it is not possible to directly analyze this ML approach, a constrained version of the ML
scheme that treats the activity pattern as a binary 0-1 vector lends itself to analysis.  
The constrained ML scheme yields a combinatorial
minimization with exponential complexity and therefore is not useful in practice. However, 
it was shown that the scaling law for successful detection of the activity pattern 
was the same as found for the NNLS estimator. 
Therefore it was conjectured that our original low-complexity ``relaxed'' ML algorithm achieves the same
scaling law.  \footnote{The analysis of the constrained ML scheme was also presented in
\cite{Kha2017} but it was based on a withdrawn RIP result \cite{Kha2019}, which has been successively fixed in our analysis in \cite{Fen2019d}.} 
A novel approach to the analysis of the the relaxed ML estimator based on the general asymptotic Gaussianity of 
the ML estimators  was recently presented in \cite{Che-Asilomar2019}. This analysis substantially corroborates 
our conjecture and provides extensive numerical evidence that  
our ML estimator can identify effectively $K_a \approx O(\dimPilots^2)$ (up to logarithmic terms), in the sense that within this regime
and for sufficiently large number of antennas $M$ the estimation error is very small with high probability. 

It is evident that the AD problem and the random access problem are related.  In fact, one can
immediately obtain a random access scheme from an AD scheme as follows: assign to each user a unique
set of pilot signature sequences (codewords), such that a user, when active, will transmit the
signature corresponding to its information message.  Since the number of pilot signatures is $K_{\rm
tot} \gg K_a$, this scheme involves only an expansion of the number of total users from $K_{\rm
tot}$ to $K'_{\rm tot} = K_{\rm tot} 2^B$ where $B$ is the number of per-message information bits.
This idea was recently presented in \cite{Sen2017}, where the MMV-AMP detector of
\cite{Kim2012,liu2017massive, Che2018} was used at the receiver side.  While conceptually simple,
this approach has two major drawbacks: 1) even for relatively small information packets (e.g., $B =
100$ bits), the dimension of the pilot matrix is too large for practical computational algorithms;
2) each user has a different set of pilot sequences, and therefore the scheme is not compliant with
the basic assumption of U-RA, that all users have the same codebook. 

In contrast, we show that a U-RA compliant scheme with a covariance based
decoder, in theory, an arbitrary small probability of
error is achievable at any $E_b/N_0$ provided that a sufficiently large number of base station
antennas is used, and that the sum spectral efficiency can grow as $\mathcal{O}(L\log(L))$.
In practice this is not possible since the complexity would grow exponentially.
To bypass this we consider a concatenated scheme, build upon the approach of
\cite{Ama2018}, that does not incur in the large dimension problem and is independent of the number
of ``inactive'' users.  In our scheme, the message of $B$ bits of each user is split into a sequence
of submessages of potentially different lengths.  These submessages are encoded via a tree code (the
same for each user), such that the encoded blocks have the same length of $J$ bits.  Then, each user
transmits its sequence of $J$-bits blocks in consecutive slots of $L$ dimensions, using the same $L
\times 2^J$ coding matrix (where blocks are encoded in the matrix columns).  The inner detector
uses the ML activity detection scheme and for each slot recovers the set of active columns of the
coding matrix.  These are passed to the outer tree code, which recovers each user message by
``stitching together'' the sequence of submessages as valid paths in the code tree. 
We show that with the concatenated code the
sum spectral efficiency can grow as $\mathcal{O}(L/\log L)$. 
So it is possible to transmit an arbitrary large amount of bits per signal dimension, 
provided that the number of antennas at the BS is large enough.
This can be achieved in a completely non-coherent way, i.e. it is at no point necessary to estimate the
channel matrix (small-scale fading coefficients). These properties make the presented scheme well
suited for easy deployable, low-latency, energy efficient communication in an IoT setting.

\section{System model}
\label{sec:model}
We consider a block-fading channel with 
blocks of $L$ signal dimensions over which the user channel vectors are constant.
We assume $n = S L$, for some integer $S$, such that the transmission of a codeword spans
$S$ fading blocks.
Following the problem formulation in \cite{Pol2017}, each user is given the same codebook
$\Cc = \{ \cv(m) : m \in [2^{nR}]\}$, formed by 
$2^{nR}$ codewords $\cv(m) \in \CC^n$. An unknown number $K_a$ out of $K_{\rm tot}$ total users transmits their
message over the coherence block.
\footnote{Here, as in 
\cite{Pol2017} and in \cite{Ama2018}, we assume that users are synchronized.
This assumption is not very restrictive since it is reasonable to assume that all users in the system can
listen to a common reference signal.}
Let $\mathcal{K}_a$ denote the set of active users.
The BS must then produce a list $\Lc$ of the transmitted messages $\{m_k : k\in \Kc_a\}$ (i.e., 
the messages of the active users). 
The system performance is expressed in terms of the {\em Per-User Probability of Misdetection}, defined as the average fraction of 
transmitted messages not contained in the list, i.e., 
\beq
p_{md} = \frac{1}{K_a}\sum_{k \in \mathcal{K}_a} \PP(m_k \notin \mathcal{L}),
\label{eq:ura_pmd}
\eeq
and the {\em Probability of False-Alarm}, defined as the average fraction of decoded messages that 
were indeed not sent, i.e., 
\beq
p_{fa} = \frac{|\mathcal{L}\setminus \{ m_k : k \in \Kc_a \}|}{|\mathcal{L}|}.
\label{eq:ura_pfa}
\eeq
The size of the list is also an outcome of the decoding algorithm, and therefore it is a random variable. 
Notice that in this problem formulation the number of total users $K_{\rm tot}$ is completely irrelevant,
as long as it is much larger than the range of possible active user set sizes $K_a$
(e.g., we may consider $K_{\rm tot} = \infty$). 
Letting the average energy per symbol of the codebook $\Cc$ be denoted by $E_s = \frac{1}{n 2^{nR}} \sum_{m=1}^{2^{nR}}  \|\cv(m)\|_2^2$,
the received signal can be re-normalized such that the AWGN per-component variance is $\sigma^2 = N_0/E_s$ and the received energy per code symbol is 1. 
Furthermore, as customary in coded systems, 
we express energy efficiency in terms of the standard quantity $E_b/N_0 :=  \frac{E_s}{R N_0}$. 

For now assume $S=1$, i.e. each user transmits his codeword in a single block of length $L$.
Further fix $J=LR$ and let $\Am \in \CC^{L \times 2^J} = [\av_1,...,\av_{2^J}]$,
be a matrix with columns normalized such that
$\| \av_i \|_2^2 = L$. Each column of $\Am$ represents one codeword.
Let $i_k$ denote the
$J$-bit messages produced by the active users $k \in \Kc_a$, represented as integers in $[1:2^J]$, 
user $k$ simply sends the column $\av_{i_k}$ 
of the coding matrix $\Am$. The received signal at the $M$-antennas BS takes on the form 
\begin{eqnarray} 
\Ym & = & \sum_{k \in \Kc_a} \sqrt{g_k} \av_{i_k} \hv^\transp_{k}  + \Zm \nonumber \\
& = & \Am \Phim \Gm^{1/2} \Hm + \Zm
\label{matrix-channel}
\end{eqnarray}
where $\Gm = \diag(g_1, \ldots, g_{K_{\rm tot}}) \in \RR^{K_{\rm tot}\times K_{\rm tot}}$
is the diagonal matrix of deterministic but unknown LSFCs,
$\Hm \in \CC^{K_{\rm tot} \times M}$ is the matrix containing, by rows, 
the user channel vectors $\hv_k$ formed by the small-scale fading antenna coefficients (Gaussian i.i.d. entries $\sim \Cc\Nc(0,1)$), 
$\Zm \in \CC^{L \times M}$ is the matrix of AWGN samples (i.i.d. entries $\sim \Cc\Nc(0,\sigma^2)$),
and $\Phim \in \{0,1\}^{2^J \times K_{\rm tot}}$ is a binary selection matrix where
for each $k \in \Kc_a$ the corresponding column $\Phim_{:,k}$ is all-zero but a single one in position $i_k$, 
and for all $k \in \Kc_{\rm tot} \setminus \Kc_a$ the corresponding column $\Phim_{:,k}$ contains all zeros. 

In line with the classical massive MIMO setting \cite{Mar2016},
we assume for simplicity an independent Rayleigh fading model, such that the channel vectors
$\{\bfh_k: k \in \clK_\text{tot}\}$ are independent from each other and are  spatially white (i.e.,
uncorrelated along the antennas), that is,  $\bfh_k \sim \cg(0, \bfI_M)$.

Let's focus on the matrix $\Xm =  \Phim \Gm^{1/2} \Hm$ of dimension $2^J \times M$. 
The $r$-th row of such matrix is given by 
\beq
\Xm_{r,:} = \sum_{k\in\mathcal{K}_a}\sqrt{g_k} \phi_{r,k} \hv^\transp_k, 
\eeq
where $\phi_{r,k}$ is the $(r,k)$-th element of  $\Phim$, equal to one if $r = i_k$ and zero otherwise. 
It follows that $\Xm_{r,:}$ is Gaussian with i.i.d. entries $\sim \Cc\Nc\left (0, \sum_{k \in \Kc_a} g_k \phi_{r,k} \right )$. 
Since the messages are uniformly distributed over $[1:2^J]$ and statistically independent across the users,
the probability that $\Xm_{r,:} $ is identically zero is given by $(1 - 2^{-J})^{K_a}$. 
Hence, for $2^J$ significantly larger than $K_a$, the matrix $\Xm$ is row-sparse. 

In order to show the equivalence to the MMV joint-sparse-support-recovery (JSSR) problem we define the modified LSFC-activity coefficients 
$\gamma_r := \sum_{k\in\mathcal{K}_a} g_k \phi_{r,k}$ and
$\Gammam = \text{diag}(\gamma_1,...,\gamma_{2^J})$. Then, (\ref{matrix-channel}) can be written as
\beq
\Ym = \Am \Gammam^{1/2} \widetilde{\Hm} + \Zm,
\label{eq:matrix_model2}
\eeq
where $\widetilde{\Hm} \in \CC^{2^J \times M}$ with i.i.d. elements $\sim \Cc\Nc(0,1)$.  
Notice that in (\ref{eq:matrix_model2}) the number of total users $K_{\rm tot}$ plays no role. 
In fact, none of the matrices involved in (\ref{eq:matrix_model2}) depends on $K_{\rm tot}$. 

The task of the decoder at the BS is to identify the non-zero elements of the modified active LSFC  pattern $\gammav$, the vector of 
diagonal coefficients of $\Gammam$. 
The active (non-zero) elements correspond to the indices of the transmitted messages. 
Notice that even if two or more users choose the same submessage, the corresponding modified LSFC
$\gamma_r$ is positive since it corresponds to the sum of the signal powers.

\section{Covariance based support recovery}

The covariance based approach to the JSSR problem \eqref{eq:matrix_model2} is based on the observation that the columns $\yv_i$
of $\Ym$, $i=1,...,M$
are Gaussian iid $\sim \mathcal{N}(0,\Sigmam_\yv)$ with covariance matrix 
\beq
    \Sigmam_\yv = \Am \Gammam \Am^\herm + \sigma^2\Id_L.
\eeq
We then attempt to recover $\Gammam = \diag(\gammav)$ from the observed empirical covariance matrix
\beq
\widehat{\Sigmam}_\yv := \frac{1}{M}\Ym\Ym^\herm.
\label{eq:samp_cov}
\eeq
For that we consider two estimators, the first is a  maximum likelihood (ML) estimator that considers the activity vector $\gammav$ as a deterministic unknown 
non-negative vector 
and the second is based on non-negative-least-squares (NNLS). 
While the former will be used in practice due
to its consistently superior performance, the latter is more suitable for giving theoretical achievability guarantees.
Both estimators can efficiently be calculated with an iterative componentwise optimization algorithm. 
For the scaling law analysis of the estimation error if the ML estimator 
versus the parameters $L, M$ and $K_a$, please refer to the 
comments in Section \ref{intro}. 

\subsection{Maximum Likelihood}   \label{ML-sect}

Let us first consider the Maximum Likelihood (ML) estimator of $\gammam$.
We introduce the log-likelihood cost function
\begin{align}
    \label{eq:likelihood_function}
f(\gammam)&:=-\frac{1}{M}\log p(\bfY| \gammam)\stackrel{(a)}{=}-\frac{1}{M}\sum_{i=1}^M \log p(\bfY_{:,i}| \gammam)\\
&=  \log | \bfA \Gammam \bfA^\herm+ \sigma^2 \bfI_{\dimPilots}| + \tr\left ( \Big( \bfA \Gammam \bfA^\herm+ \sigma^2 \bfI_{\dimPilots}\Big ) ^{-1} \widehat{\Sigmam}_\bfy \right),\label{eq:ML_cost}
\end{align} 
where $(a)$ follows from the fact that the columns of $\bfY$ are i.i.d.
(due to the spatially white user channel vectors),
and where $\widehat{\Sigmam}_\bfy$ denotes the sample covariance matrix of the columns of $\bfY$ as in \eqref{eq:samp_cov}.
We define the ML estimator as
\begin{align}
    \gammam^*_\text{ML} = \argmin_{\gammam \in \bR_+^{K_\text{tot}}} f(\gammam).\label{a_ML}
\end{align}
Note that the set over that we search for the minimum is larger than the actual non-convex
set of permitted signals $\{\gammav \in \bR_+^{K_{\rm tot}}: \|\gammam\|_0 \leq K_a\}$. In that
sense \eqref{a_ML} is a relaxed version of the actual (constrained) ML estimator.
Nonetheless, the constrained ML estimator requires optimization over an exponentially large
number of support combinations, which is in general infeasible, while the relaxed version shows to be efficiently
computable. Furthermore, \eqref{a_ML} does not require any prior knowledge of $K_a$. 

\subsection{Non-Negative Least Squares}  \label{NNLS-sect}

\label{sec:nnls}
For
\beq
\Sigmam(\gammam) := \bfA \Gammam\bfA^\herm + \sigma^2\Id_{\dimPilots}
\label{eq:Sigma_def}
\eeq 
we define the NNLS estimator as
\begin{align}\label{eq_nnls}
    \gammam^*_\text{NNLS} =\argmin_{\gammam \in \bR_+^K} \|\Sigmam(\gammam) - \widehat{\Sigmam}_\bfy\|^2_{\sfF}.
\end{align}
Let us introduce the matrix $\bA\in\CC^{\dimPilots^2\times K_\text{tot}}$, whose $k$-th column is defined by:
\beq
\bA_{:,k} := \text{vec}(\av_k\av_k^\herm).
    \label{eq:bA_definition}
\eeq
and let $\wv = \vec(\widehat{\Sigmam}_\bfy - \sigma^2\Id_{\dimPilots})$ 
denote the
$\dimPilots^2 \times 1$ vector obtained by stacking the columns of
$\widehat{\Sigmam}_\bfy- \sigma^2\Id_{\dimPilots}$.
Then, we can write \eqref{eq_nnls} in the convenient form
\begin{align}\label{eq_nnls_vec}
  \gammam^*=\argmin_{\gammam \in \bR_+^K} \|\bA \gammam -\wv\|^2_2,
\end{align}
as a linear \textit{least squares} problem with non-negativity constraint,
known as \textit{non-negative least squares} (NNLS).
Such an algorithm was proposed for the activity detection problem in \cite{docomo}.

NNLS has a special property, as discussed for
example in \cite{slawski2013non} and referred to as the $\mathcal{M}^+$-criterion in \cite{kueng2016robust}, 
which makes it particularly suitable for recovering sparse vectors:
If  the row span of $\bA$ intersects the positive orthant, NNLS
implicitly also performs $\ell_1$-regularization.  

\subsection{Iterative Algorithm}  \label{iter-algo}
The presented estimators can both be found using
an {\em iterative componentwise  minimization algorithm}: 
Starting from an initial point $\gammam$, 
at each step of the algorithm we minimize the objective function $f(\gammam)$
with respect to only one of its coordinates $\gamma_k$. 
Of course, the component update step is different in the case of ML and in the case of NNLS. 
The derivation of the update rules can be found in \cite{Fen2019d},
we summarize them in Algorithm \ref{tab:ML_coord}.
Variants of the algorithm may differ in the way the initial point is chosen and in the way
the components are chosen for update. We can also include the noise variance $\sigma^2$ as an additional 
optimization parameter and estimate it along $\gammam$. 

\begin{algorithm}[t]
	\caption{Joint support recovery via Coordinate-wise Optimization } %
	\label{tab:ML_coord} 
{\small
	\begin{algorithmic}[1]
		\State {\bf Input:} The sample covariance matrix $\widehat{\Sigmam}_\bfy=\frac{1}{M} \bfY \bfY^\herm$ of the $\dimPilots \times M$ matrix of samples $\bfY$.
		
		\State {\bf Input:} The LSFCs of $K_\text{tot}$ users $(g_1, \dots, g_{K_\text{tot}})$ if available.
		
		\vspace{1mm}

		\State {\bf Initialize:} $\Sigmam=\sigma^2 \bfI_{\dimPilots}$, $\gammam={\bf 0}$.
		\vspace{1mm}
		
		\For { $i=1,2, \dots$}
		\State {Select an index $k \in [K_\text{tot}]$ corresponding to the $k$-th component of  $\gammam=(\gamma_1, \dots, \gamma_{K_\text{tot}})^\transp$  randomly or according to a specific schedule.}
		\vspace{2mm}
		\State {\bf ML:} Set $d^*= \max \Big \{\frac{ \bfa_k^\herm \Sigmam^{-1} \widehat{\Sigmam}_\bfy \Sigmam^{-1} \bfa_k -  \bfa_k^\herm \Sigmam^{-1}\bfa_k }{(\bfa_k^\herm \Sigmam^{-1}\bfa_k )^2}, -\gamma_{k} \Big \}$.
		
		\State {\bf NNLS:} Set $d^*=\max \Big \{  \frac{\bfa_k^\herm (\widehat{\Sigmam}_\bfy - \Sigmam ) \bfa_k}{\|\bfa_k\|_2^4}, -\gamma_{k} \}$. 
		
		\vspace{1mm}
		
        \label{alg:update_step}
		
		\State Update $\gamma_{k} \leftarrow \gamma_{k}+ d^*$.
		\State Update 
        $\Sigmam^{-1}
\leftarrow \Sigmam^{-1} -
\frac{d^* \Sigmam^{-1} \bfa_k \bfa_k^\herm \Sigmam^{-1}}
{1+ d^* \bfa_k^\herm \Sigmam^{-1} \bfa_k}$
		\EndFor
		\State {\bf Output:}  The resulting estimate $\gammam$. 
	\end{algorithmic}}
\end{algorithm}

\subsection{Asymptotic scaling}  
\label{sec:analysis}

In this section we discuss the performance of the NNLS estimator in a single slot ($S=1$).
For the sake of simplicity, in the discussion of this section we assume $g_k = 1$ for all $k$.
In this case, the SNR $E_s/N_0$ is also the SNR at the receiver, for each individual (active) user and
$\|\gammav\|_1/\sigma^2 = K_a E_s/N_0$.
It is shown in \cite[Corollary 2]{Fen2019c} that
\begin{align}
    \frac{\lVert \gammav - \gammav^*\rVert_1}{\|\gammav\|_1 } \leq
    c   \left(1+\left(K_a\frac{E_s}{N_0}\right)^{-1}\right)\sqrt{\frac{K_a}{M}}   
\label{eq:gamma_perf}
\end{align}
holds with high probability provided that 
\beq
K_a = \mathcal{O}(L^2/\log^2(e2^J/L^2))
\label{eq:phase_transition}
\eeq
where $c>0$ is some universal constant and $\gammav^*$ denotes the estimate of
$\gammav$ by the NNLS algorithm (see section \ref{sec:nnls}).
Numerical results \cite{Fen2019d} suggest that the reconstruction error of the ML algorithm
is at least as good as that of NNLS (in practice it is {\em much better}). 
This bound is indeed very conservative.
Nevertheless, this is enough to give achievable scaling laws for the probability of
error of the decoder.  It follows from \eqref{eq:gamma_perf} that $\frac{\lVert \gammav - \gammav^*\rVert_1}{\|\gammav\|_1 }\to 0$
for $(M,K_a, \frac{E_s}{N_0}) \to (\infty, \infty, 0)$ as long as 
\beq
\frac{K_a( 1 + (K_aE_s/N_0)^{-1})^2}{M} = o(1),
\label{eq:scaling}
\eeq
which is satisfied if $M$ grows as
\beq
M = \max(K_a,(E_s/N_0)^{-1})^{\kappa}
\label{eq:M_lower}  
\eeq
for some $\kappa >1$.
Assuming that $J$ scales such that $2^J = \delta L^2$ for some fixed $\delta\geq 1$,
i.e. $J=\mathcal{O}(\log L)$, then the scaling condition \eqref{eq:phase_transition} becomes 
$K_a = \mathcal{O}(L^2)$ and we can conclude that the recovery error vanishes for sum spectral efficiencies
up to
\beq
\frac{K_aJ}{L} = \mathcal{O}(L\log L).
\eeq
This shows that, in principles, we can achieve a total spectral efficiency that grows without bound, by encoding
over larger and larger blocks of dimension $L$, as long as the number of messages per user
and the number of active users both grow
proportionally to $L^2$, when the number of BS antennas scales as in (\ref{eq:M_lower}),
this system achieves a sum spectral efficiency that
grows with $L \log(L)$ and an error  probability as small as desired, for any given $E_b/N_0 > 0$. 
Of course, in this regime the rate per active user vanishes as $\log(L)/L$. 
\section{Reducing complexity via concatenated coding}
\label{sec:coding_concatenated}
In practice it is not feasible to transmit even small messages (e.g. $J\sim100$) within one coherence block ($S=1$),
because the number of columns in the coding matrix $\Am$ grows exponentially in $J$. 
Let each user transmit his message over a \emph{frame} of $S$ fading blocks and within each block use the code
described in section \ref{sec:model} as \emph{inner} code with the ML decoder as inner decoder.

We follow the concatenated coding scheme approach of \cite{Ama2018},
suitably adapted to our case.  Let $B = nR$ denote the number of bits per user message. 
For some suitable integers $S \geq 1$ and $J > 0$, we divide the $B$-bit message into blocks of size 
$b_1, b_2, \ldots, b_S$ such that $\sum_s b_s = B$ and such that
$b_1 =  J$ and $b_s < J$ for all $s = 2, \ldots, S$. 
Each subblock $s = 2, 3, \ldots, S$ is augmented to size $J$ by appending $p_s = J - b_s$ parity bits,  
obtained using pseudo-random linear combinations of the information bits of the previous blocks 
$s' < s$. Therefore, there is a one-to-one association between the set of all sequences of coded blocks and
the paths of a tree of depth $S$. The pseudo-random parity-check equations generating the parity bits 
are identical for all users, i.e., each user makes use exactly of the same outer {\em tree code}.
For more details on the outer coding scheme, please refer to  \cite{Ama2018}. 

Given $J$ and the slot length $L$, the inner code is used to transmit in sequence the 
$S$ (outer-encoded) blocks forming a frame. Let $\Am$ be the coding matrix as defined in section \ref{sec:model}
Each column of $\Am$ now represents one inner codeword.
Letting $i_k(1), \ldots, i_k(S)$ denote the sequence of $S$ (outer-)encoded 
$J$-bit messages produced by the outer encoder of active user $k \in \Kc_a$. The user $k$ now simply sends in
sequence, over consecutive slots of length $L$, the columns $\av_{i_k(1)},\av_{i_k(2)},...,\av_{i_k(S)}$
of the coding matrix $\Am$. As described in section \ref{sec:model}, the inner decoding problem is equivalent to
the AD problem \eqref{eq:matrix_model2}.
For each subslot $s$, let $\widehat{\gammav}[s] = (\widehat{\gamma}_1[s], \ldots, \widehat{\gamma}_{2^J}[s])^\trasp$ 
denote the ML estimate of $\gammav$ in subslot $s$ obtained by the inner decoder. 
Then, the list of active messages at subslot $s$ is defined as
\begin{equation}
\label{support-detection} 
\Sc_s = \left \{ r \in [2^J] : \widehat{\gamma}_r[s] \geq \nu_s \right \}, 
\end{equation}
where $\nu_1, \ldots, \nu_S$ are suitable pre-defined thresholds.  
Let $\Sc_1, \Sc_2, \ldots, \Sc_S$ the sequence of lists of active subblock 
messages. Since the subblocks contain parity bits with parity profile $\{0,p_2, \ldots, p_S\}$, 
not all message sequences in $\Sc_1 \times \Sc_2 \times \cdots \times \Sc_S$ are possible. 
The role of the outer decoder is to identify all possible message sequences, i.e., those corresponding to
paths in the tree of the outer tree code \cite{Ama2018}.  
The output list $\Lc$ is initialized as an empty list. Starting from $s=1$ and proceeding in order, the decoder converts the integer indices 
$\Sc_s$ back to their binary representation, separates data and parity bits, computes the parity 
checks for all the combinations with messages from the list $\Lc$ and extends only the paths 
in the tree which fulfill the parity checks.  A precise analysis of the error probability of such a decoder
and its complexity in terms of surviving paths in the list is given in \cite{Ama2018}.

The performance of the concatenated system
is demonstrated via simulations in section \ref{sec:numerics}.

\subsection{Asymptotic analysis - Outer code}
\label{sec:analysis2}

We define the support $\rhov[s]$ of the estimated $\widehat{\gammav}[s]$ as a binary vector
whose $r$-th element is equal to 1 if  $\widehat{\gamma}_r[s] \geq \nu_s$ and to zero otherwise. 
In the case of error-free support recovery, 
$\rhov[s]$  can be interpreted as the output of a vector ``OR'' multiple access channel (OR-MAC) where the inputs are the binary columns of the activity matrix 
$\Phim[s]$ and the output is given by 
\beq
\rhov[s] =  \bigvee_{k \in \Kc_a} \Phim_{:,k}[s], 
\eeq
where $\bigvee$ denotes the component-wise binary OR operation.  
The logical ``OR'' arises from the fact that if the same submessage is selected by 
multiple users, it will shows up as ``active'' at the output of the ``activity-detection'' inner decoder since the signal energy adds up
(as discussed before).

\subsubsection{Achievability}
The analysis in \cite{Ama2018} shows that the error probability of the outer code goes to zero
in the so called logarithmic regime with constant outer rate, i.e. for $K_a,J\to \infty$
as $J = \alpha \log_2 K_a$ and $B = SR_\text{out}J$
\footnote{We deviate slightly from the notation in \cite{Ama2018}, where the scaling parameter
$\alpha'$ is defined by
$B = \alpha'\log_2 K_a$ and the number of subslots is considered to be constant. It is apparent that
those definitions are connected by $\alpha' = SR_\text{out}\alpha$.}
if the number of parity bits $P$ is chosen as (\cite[Theorem 5 and 6]{Ama2018})
\begin{enumerate}
    \item $P = (S+\delta - 1)\log_2 K_a$ for some constant $\delta>0$ if all the parity bits
        are allocated in the last slots.
    \item $P = c(S-1)\log_2 K_a$ for some constant $c>1$ if the parity bits are allocated evenly
        at the end of each subslot except for the first.
\end{enumerate}
In the first case the complexity scales like $\mathcal{O}(K_a^{R_\text{out}S}\log K_a)$,
since there is no pruning in the first $R_\text{out}S$ subslots,
while in the second case the complexity scales linearly with $S$
like $\mathcal{O}(SK_a\log K_a)$. The corresponding outer rates are
\beq
\begin{split}
    R_\text{out} &= B/(B+P)\\
                 &= 1 - P/(B+P)\\
                 &= 1 - P/(SJ)\\
                 &= 1 - \frac{S+\delta - 1}{S\alpha}\\
                 &= 1 - \frac{1}{\alpha}  + \frac{1}{S}\frac{\delta-1}{\alpha}
\end{split}
\eeq
for the case of all parity bits in the last sections and
\beq
\begin{split}
R_\text{out} &= 1 - \frac{c(S-1)}{S\alpha}\\
             &= 1 - \frac{c}{\alpha}  - \frac{c}{S\alpha}
\end{split}
\eeq
for the case of equally distributed parity bits. In the limit $S\to\infty$ the achievable rates are
therefore $R_\text{out} = 1 - 1/\alpha$ and $R_\text{out} = 1 - c/\alpha$ respectively.

\subsubsection{Converse}
The output entropy of the vector OR-MAC of dimension $2^J$ is bounded by the entropy of 
$2^J$ scalar OR-MACs. The marginal distribution of the entries of $\rhov[s]$  is Bernoulli with 
$\PP(\rho_r[s] = 0) = (1 - 2^{-J})^{K_a}$. Hence, we have
\begin{equation} \label{outputH}
H(\rho[s]) \leq 2^J \Hc_2 ( (1 - 2^{-J})^{K_a}). 
\end{equation}
We stay in the logarithmic scaling regime, introduced in the previous sections,
i.e. we fix $J = \alpha \log_2 K_a$ for some $\alpha >1$ and consider the limit $K_a,J\to\infty$.
In this regime
$K_a/2^J = K_a^{-(\alpha - 1)}\to 0$ and we have
$1 - (1-2^{-J})^{K_a} = K_a/2^J + \mathcal{O}((K_a/2^J)^2) \to 0$.
This gives that
\beq
2^J\Hc_2( (1 - 2^{-J})^{K_a}) \to K_a(J-\log_2 K_a) = (\alpha - 1)K_a\log_2 K_a.
\eeq
Since all users make use of the same code
we have that the number of information bits sent by the $K_a$ active users over a slot is  
$B_\text{sum} = K_a J R_\text{out}$.
Therefore, in order to hope for small probability of error a necessary condition is
\beq
    K_a J R_\text{out} \leq 2^J\mathcal{H}_2((1-1/2^J)^{K_a}).
    \label{eq:sumrate}
\eeq
So the outer rate is limited by 
\beq
    R_\text{out} \leq (\alpha-1)\frac{\log_2 K_a}{J} = 1 - \frac{1}{\alpha}.
\eeq
We have shown in the previous section that this outer rate can be achieved in the limit of infinite subslots
$S \to \infty$ by the described outer tree
code at the cost of a decoding complexity of at least $\mathcal{O}(K_a^{R_\text{out}S})$
or up to a constant factor
$\Delta R_\text{out} = (c-1)/\alpha$ for some $c>1$ with a complexity of $\mathcal{O}(SK_a\log K_a)$.
This is a noteworthy results on its own, since it is a priori not clear, whether the bound 
\eqref{eq:sumrate} is achievable by an unsourced random access scheme, i.e. each user using
the same codebook.\\
The resulting achievable sum spectral efficiency can be calculated as in section
\ref{sec:analysis} with a subtle
but important difference, since the results on the outer code are valid only in the logarithmic regime
$J = \alpha\log_2 K_a$, i.e. $2^J = K_a^\alpha$ for $\alpha >1$.
According to \cite[Corollary 2]{Fen2019d} the error probability of
the inner code vanishes if the number of active users
scale no faster then $K_a = \mathcal{O}(L^2/\log^2(e2^J/L^2))$.
Using the scaling condition $J = \alpha \log_2 K_a$ and that $K_a \leq L^2$,
this implies that in the logarithmic regime the error probability of the inner code vanishes
if the number of active users scales as $K_a = \mathcal{O}(L^2/\log^2(L))$.
This gives a sum spectral efficiency
of 
\beq
\frac{K_aR_\text{out}J}{L} = \mathcal{O}\left(\frac{K_a\log K_a}{L}\right)
= \mathcal{O}\left(\frac{L}{\log L}\right).
\eeq
The order of this sum spectral efficiency is, by a factor $\log^2 L$, smaller then the one we calculated
in section \ref{sec:analysis}.
This is because the order of supported active users is smaller by exactly the same
$\log^2 L$ factor.
In section \ref{sec:analysis} we assumed that $J$ scales as $2^J = \delta L^2 = \mathcal{O}(K_a)$
for some $\delta>1$,
so that the ratio $K_a/2^J$ remains constant. It is not clear from the analysis in \cite{Ama2018},
whether the probability of error of the outer tree code would vanish in the regime.
We can get a converse by evaluating the entropy bound \eqref{eq:sumrate}.
Let $2^J = \delta K_a$ with $\delta>1$, then 
$(1-2^{-J})^{K_a} = (1-\delta/K_a)^{K_a}\xrightarrow[K_a\to\infty]{} \exp(-\delta)$.
Therefore the binary entropy $\Hc_2((1-2^{-J})^{K_a})$ remains a constant in the limit
$J,K_a \to \infty$ and we get that 
\beq
K_aR_\text{out}J\leq \delta K_a \Hc_2(\exp(-\delta)).
\eeq
This shows that $R_\text{out} \to 0$
in the limit $J,K_a \to \infty$ is the best achievable asymptotic per-user outer rate,
but the outer sum rate $K_aR_\text{out}J$ is proportional to $K_a$.
The resulting sum spectral efficiencies scale as
\beq
\frac{K_aR_\text{out}J}{L} = \mathcal{O}\left(\frac{K_a}{L}\right) = \mathcal{O}(L).
\eeq
This means it could be possible to increase the achievable sum spectral efficiencies by a factor of $\log L$
by using an outer code that is able to achieve the entropy bound \eqref{eq:sumrate}
in the regime $2^J = \delta K_a$. It is not clear though whether the code of \cite{Ama2018} or some other code can achieve
this.

\section{Simulations}
\label{sec:numerics}

The outer decoder requires a hard decision on the support of the estimated $\widehat{\gammav}[s]$. 
When $K_a$ is known, one approach consists of selecting the $K_a + \Delta$ largest entries in each section, where $\Delta \geq 0$
can be adjusted to balance between false alarm and misdetection in the outer channel.
However, the knowledge of $K_a$ is a very restrictive assumption in such type of systems. 
An alternative approach, which does not require this knowledge,
consists of fixing a sequence of thresholds $\{\nu_s : s \in [S]\}$ and let $\rhov[s]$ to be the binary vector of dimension $2^J$ 
with elements equal to 1 for all components of $\widehat{\gammav}[s]$ above threshold $\nu_s$. 
By choosing the thresholds, we can balance between missed detections and false alarms. 
Furthermore, we may consider the use of a  non-uniform decaying power allocation across the 
slots as described in \cite{Fen2019c}.

\begin{figure}
    \centering
    \includegraphics[width=0.8\linewidth]{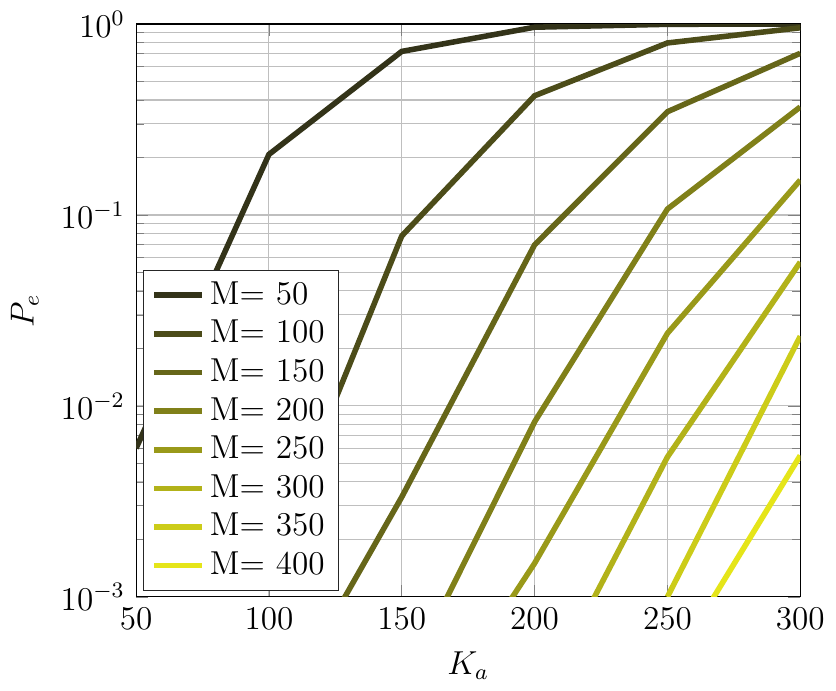}
    \captionof{figure}{ Error probability ($P_e = p_\text{md} + p_\text{fa}$) as a function of the number of
    active users for different numbers of receive antennas.
    $E_b/N_0=$ 0 dB, $L=100$, $n=3200$, $b = 96$ bits, $S = 32, J=12$. }
    \label{fig:sim}
\end{figure}

For the simulations in \figref{fig:sim} we choose $B = 96$ bits as payload size for each user, a
frame of choose $S = 32$ slots of $L = 100$ dimensions per slot, yielding an overall block length $n
= 3200$. Choosing the binary subblock length $J = 12$, the inner coding matrix $\Am$ has dimension
$100 \times 4096$ and therefore is still quite manageable. We choose the columns of $\Am$ uniformly
i.i.d. from the sphere of radius $\sqrt{L}$. Notice also that if one wishes to send the same payload
message using the piggybacking scheme of  \cite{Lar2012,Sen2017}, each user should make use of
$2^{96}$ columns, which is totally impractical. 

\begin{figure}
   \centering
   \subfloat[$K_a = 300$]{\includegraphics[height=6cm]{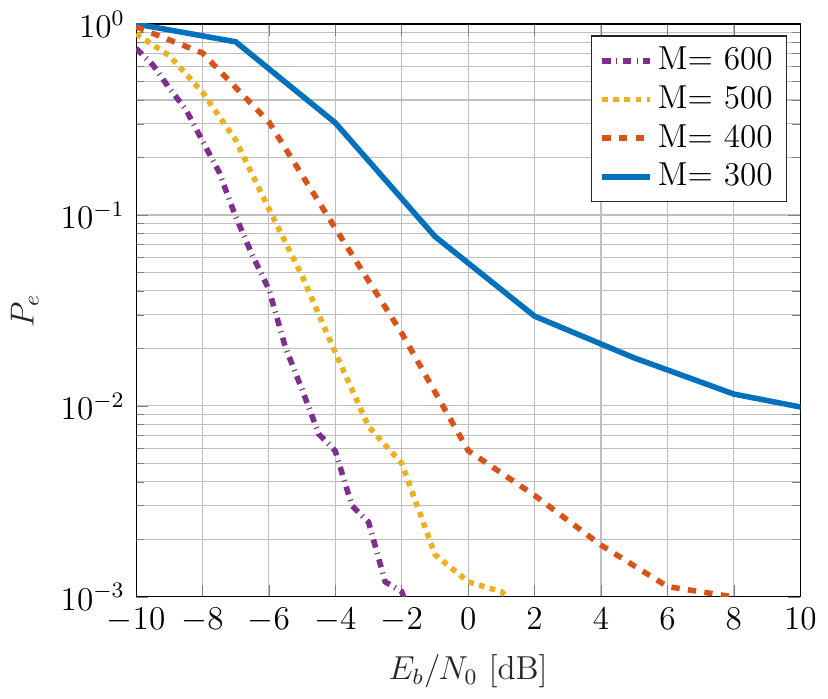}}\quad
   \subfloat[$M=300$]{\includegraphics[height=6cm]{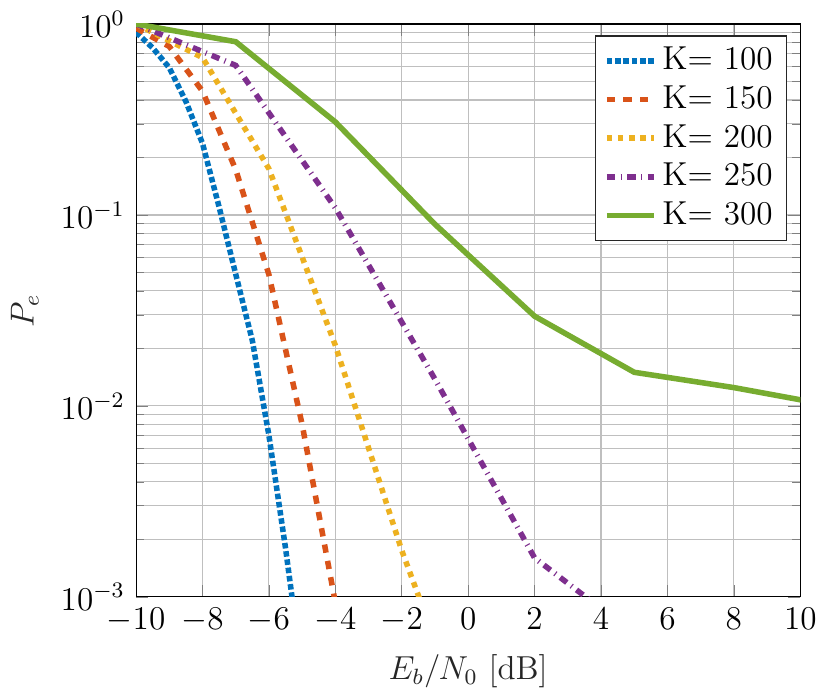}}
  \caption{Error probability ($P_e = p_\text{md} + p_\text{fa}$) as a function of $E_b/N_0$.
    $L=100$, $n=3200$, $b = 96$ bits, $S = 32, J=12$. }
    \label{fig:pe_ebn0}
\end{figure}

For the outer code, we choose the following parity profile $p = [0,9,9, \ldots ,9,12,12,12]$,
yielding an outer coding rate $R_{out} = 0.25$ information bits per binary symbol. 
All large scale fading coefficients are fixed to $g_k \equiv 1$.
In \figref{fig:sim} we fix $N_0 = 1$ and choose the transmit power (energy per symbol),
such that $E_b/N_0 = 0$dB and plot the sum of the
two types of error probabilities
$P_e = p_\text{md} + p_\text{fa}$ (see \eqref{eq:ura_pmd} and \eqref{eq:ura_pfa})
as a function of the number of active users for
different numbers of receive antennas $M$.  
\figref{fig:pe_ebn0} shows how $P_e$ falls as a function of $E_b/N_0$ for different values of $K_a$ and $M$.
Table \ref{tab:ebn0} summarizes the required values of $E_b/N_0$ to achieve a total error probability
$P_e < 0.05$.
Notice that this corresponds to a total spectral efficiency $\mu = \frac{12}{100} \times 0.25 \times 300 = 9$ bit per channel 
use, which is  significantly larger than today's LTE cellular systems (in terms of bit/s/Hz per sector) and definitely
much larger than IoT-driven schemes such as  LoRA \cite{Cen2016,Ban2016}. 
According to the random coding bound of \cite{Pol2017} this is impossible to achieve for the scalar Gaussian channel 
(only one receive antenna), even with coherent detection and roughly five times 
smaller spectral efficiency then here.
This shows also quantitatively that the non-coherent massive MIMO channel is very attractive for 
U-RA, since it preserves the same attractive characteristics of U-RA as in the 
non-fading Gaussian model of \cite{Pol2017} (users transmit without any pre-negotiation, and no use of pilot symbols
is needed), while the total spectral efficiency can be made as large as desired simply by increasing the number of receiver 
antennas. 

\begin{table}
    \centering
    \subfloat[$K_a=300$]{
    \begin{tabular}{|c|c|c|c|c|}
        \hline
        $M$ & 300 & 400 & 500 & 600 \\ \hline
        $E_b/N_0$ [dB]& 0.4 & -3.1 & -5.0 & -6.2 \\ \hline
    \end{tabular}
}\quad
    \subfloat[$M=300$]{
    \begin{tabular}{|c|c|c|c|c|c|}
        \hline
        $K_a$ & 100 & 150 & 200 & 250 & 300 \\ \hline
        $E_b/N_0$ [dB]& -7.0 & -6.0 & -4.8 & -2.9 & 0.6\\ \hline
    \end{tabular}
}
    \caption{Required $E_b/N_0$ to achieve a total error probability $P_e < 0.05$
    with $L=100$, $n=3200$, $b = 96$ bits, $S = 32, J=12$.}
    \label{tab:ebn0}
\end{table}

\section{Conclusion}

In this paper we studied the problem of unsourced random access,
a special type of grant-free random access, for the block-fading channel with a massive MIMO BS.
We showed that an arbitrarily fixed probability of error can be achieved
at any $E_b/N_0$ for sufficiently large number of antennas, and a total spectral efficiency that grows as $O(L \log L)$,
where $L$ is the code block length, can be achieved.
Such one-shot scheme is conceptually nice but not suited for typical practical applications with message payload
of the order of $B \approx 100$ bits, since it would require a codebook matrix with $2^B$ columns.
Hence, we have also considered the application of the concatenated approach pioneered in \cite{Ama2018},
where the message is broken into a sequence of smaller blocks
and the activity detection scheme is applied as an inner encoding/decoding stage at each block,
while an outer tree code takes care of ``stitching together'' the sequence of decoded submessages over the blocks.
We show that the concatenated coding scheme can achieve a total spectral efficiency of $\mathcal{O}(L/\log L)$.
Numerical simulations show the effectiveness of the proposed method.
It should be noticed that these schemes are completely non-coherent,
i.e., the receiver never tries to estimate the massive MIMO channel matrix of complex fading coefficients.
Therefore, the scheme pays no hidden penalty in terms of pilot symbol overhead,
often connected with the assumption of ideal coherent reception,
i.e., channel state information known to the receiver.

\section*{Acknowledgement} 
P.J. is supported by DFG grant JU 2795/3.

\balance 

{\small
\bibliographystyle{IEEEtran}
\bibliography{references}
}

\end{document}

%% file: symbols.tex
\usepackage{mathbbol}

\def\mindex#1{\index{#1}}

\def\sq{\hbox{\rlap{$\sqcap$}$\sqcup$}}
\def\qed{\ifmmode\sq\else{\unskip\nobreak\hfil
\penalty50\hskip1em\null\nobreak\hfil\sq
\parfillskip=0pt\finalhyphendemerits=0\endgraf}\fi\medskip}

\long\def\defbox#1{\framebox[.9\hsize][c]{\parbox{.85\hsize}{%
\parindent=0pt
\baselineskip=12pt plus .1pt      %
\parskip=6pt plus 1.5pt minus 1pt %
 #1}}}

\long\def\beginbox#1\endbox{\subsection*{}%
\hbox{\hspace{.05\hsize}\defbox{\medskip#1\bigskip}}%
\subsection*{}}

\def\endbox{}

\def\diag{{\text{diag}}}

\def\tr{\mathsf{tr}}

\newsavebox{\junk}
\savebox{\junk}[1.6mm]{\hbox{$|\!|\!|$}}

\def\argmin{\mathop{\rm arg\, min}}

\def\bA{{\mathbb A}}

\def\bR{{\mathbb R}}

\def\bfA{{\bf A}}

\def\bfI{{\bf I}}

\def\bfY{{\bf Y}}

\def\bfa{{\bf a}}

\def\bfh{{\bf h}}

\def\bfy{{\bf y}}

\def\sfF{{\sf F}}

\def\sfH{{\sf H}}

\def\bfmath#1{{\mathchoice{\mbox{\boldmath$#1$}}%
{\mbox{\boldmath$#1$}}%
{\mbox{\boldmath$\scriptstyle#1$}}%
{\mbox{\boldmath$\scriptscriptstyle#1$}}}}

\def\bfmY{\bfmath{Y}}

\def\bfmhhaY{\bfmath{\hhaY}} %
\def\bfmhhaY{\hbox to 0pt{$\widehat{\bfmY}$\hss}\widehat{\phantom{\raise 1.25pt\hbox{$\bfmY$}}}}

\def\til={{\widetilde =}}

\def\clC{{\cal C}}

\def\clK{{\cal K}}

\def\clN{{\cal N}}

 \def\FRAC#1#2#3{\genfrac{}{}{}{#1}{#2}{#3}}

\def\ddtp{{\mathchoice{\FRAC{1}{d^{\hbox to 2pt{\rm\tiny +\hss}}}{dt}}%
{\FRAC{1}{d^{\hbox to 2pt{\rm\tiny +\hss}}}{dt}}%
{\FRAC{3}{d^{\hbox to 2pt{\rm\tiny +\hss}}}{dt}}%
{\FRAC{3}{d^{\hbox to 2pt{\rm\tiny +\hss}}}{dt}}}}

\def\average#1,#2,{{1\over #2} \sum_{#1}^{#2}}

\def\eye(#1){{\bf(#1)}\quad}

\def\eq#1/{(\ref{e:#1})}

\newcommand\figref{Figure~\ref}

\newcommand{\beqn}[1]{\notes{#1}%
\begin{eqnarray} \elabel{#1}}

\newcommand{\eeqn}{\end{eqnarray} }

\newcommand{\beq}{\begin{equation}}

\newcommand{\eeq}{\end{equation}}

\def\bdes{\begin{description}}
\def\edes{\end{description}}

\newcounter{rmnum}

\newcounter{anum}

{\end{list}}

\def\ass(#1:#2){(#1\ref{#1:#2})}

\def\ritem#1{
\item[{\sf \ass(\current_model:#1)}]
}

\newenvironment{recall-ass}[1]{%
\begin{description}
\def\current_model{#1}}{
\end{description}
}

%% file: macros.tex
\long\def\comment#1{}

\newcommand{\CC}{\mathbb{C}}
\newcommand{\PP}{\mathbb{P}}
\newcommand{\RR}{\mathbb{R}}

\newcommand{\av}{{\bf a}}

\newcommand{\cv}{{\bf c}}

\newcommand{\hv}{{\bf h}}

\newcommand{\wv}{{\bf w}}

\newcommand{\yv}{{\bf y}}

\newcommand{\Am}{{\bf A}}

\newcommand{\Gm}{{\bf G}}
\newcommand{\Hm}{{\bf H}}
\newcommand{\Id}{{\bf I}}

\newcommand{\Xm}{{\bf X}}
\newcommand{\Ym}{{\bf Y}}
\newcommand{\Zm}{{\bf Z}}

\newcommand{\Cc}{{\cal C}}

\newcommand{\Hc}{{\cal H}}

\newcommand{\Kc}{{\cal K}}
\newcommand{\Lc}{{\cal L}}

\newcommand{\Nc}{{\cal N}}

\newcommand{\Sc}{{\cal S}}

\newcommand{\gammav}{\hbox{\boldmath$\gamma$}}

\newcommand{\rhov}{\hbox{\boldmath$\rho$}}

\newcommand{\Gammam}{\boldsymbol{\Gamma}}

\newcommand{\Sigmam}{\boldsymbol{\Sigma}}
\newcommand{\Phim}{\boldsymbol{\Phi}}

\newcommand{\trasp}{{\sf T}}
\newcommand{\transp}{{\sf T}}
\renewcommand{\vec}{{\rm vec}}